\documentclass[12pt]{article}
\usepackage{graphicx}
\usepackage{xspace}
\usepackage{verbatim}

\newcommand\pubnumber{SNSN-323-63}
\newcommand\pubdate{January 15, 2020}

\def\institute{
	Deutsches Elektronen-Synchrotron (DESY) \\
	Notkestr.\,85, 22607 Hamburg, GERMANY
}

\def\Title#1{\begin{center} {\Large #1 } \end{center}}
\def\Author#1{\begin{center}{ \sc #1} \end{center}}
\def\Address#1{\begin{center}{ \it #1} \end{center}}

\newcommand\pubblock{\rightline{\begin{tabular}{l} \pubnumber\\
         \pubdate  \end{tabular}}}
\newenvironment{Abstract}{\begin{quotation}  }{\end{quotation}}
\newenvironment{Presented}{\begin{quotation} \begin{center} 
             PRESENTED AT\end{center}\bigskip 
      \begin{center}\begin{large}}{\end{large}\end{center} \end{quotation}}
\def\Acknowledgements{\bigskip  \bigskip \begin{center} \begin{large}
             \bf ACKNOWLEDGEMENTS \end{large}\end{center}}

\def\beq{\begin{equation}}
\def\eeq#1{\label{#1}\end{equation}}
\def\eeqn{\end{equation}}

\def\beqa{\begin{eqnarray}}
\def\eeqa#1{\label{#1}\end{eqnarray}}
\def\eeqan{\end{eqnarray}}

\let\bar=\overbar

\def\ie{{\it i.e.}\xspace}

\def\Dslash{\not{\hbox{\kern-4pt $D$}}}
\def\dslash{\not{\hbox{\kern-2pt $\del$}}}

\def\mt{\ensuremath{m_\mathrm{t}}\xspace}

\def\msb{{\bar{\ssstyle M \kern -1pt S}}}

\newcommand{\msbar}{\ensuremath{\mathrm{\overline{MS}}}\xspace}
\newcommand{\ttbar}{\ensuremath{\mathrm{t\bar{t}}}\xspace}
\newcommand{\fbinv}{\ensuremath{\;\mathrm{fb}^{-1}}\xspace}
\newcommand{\TeV}{\ensuremath{\,\mathrm{TeV}}\xspace}
\newcommand{\as}{\ensuremath{\alpha_\mathrm{S}}\xspace}
\newcommand{\mtt}{\ensuremath{m_\mathrm{t\bar{t}}}\xspace}
\newcommand{\stt}{\ensuremath{\sigma_{\ttbar}}\xspace}
\newcommand{\dstt}{\ensuremath{\mathrm{d}\stt/\mathrm{d}\mtt}\xspace}
\newcommand{\pt}{\ensuremath{\mathrm{p_T}}\xspace}
\newcommand{\GeV}{\ensuremath{\,\mathrm{GeV}}\xspace}
\newcommand{\mttreco}{\ensuremath{\mtt^\mathrm{reco}}\xspace}
\newcommand{\mtmc}{\ensuremath{m_\mathrm{t}^{\mathrm{MC}}}\xspace}

\newcommand{\muk}{\ensuremath{\mu_k}\xspace}

\newcommand{\mlb}{\ensuremath{m_{\ell\mathrm{b}}^{\mathrm{min}}}\xspace}

\newcommand{\mtmt}{\ensuremath{m_\mathrm{t}(m_\mathrm{t})}\xspace}
\newcommand{\MCFM} {{\textsc{MCFM}}\xspace}
\newcommand{\mtmuk}{\ensuremath{m_\mathrm{t}(\mu_k)}\xspace}
\newcommand{\chisq}{\ensuremath{\chi^2}\xspace}
\newcommand{\mtmu}{\ensuremath{m_\mathrm{t}(\mu)}\xspace}
\newcommand{\muref}{\ensuremath{\mu_\mathrm{ref}}\xspace}
\newcommand{\mtot}{\ensuremath{\mt^\mathrm{incl}(\mt)}\xspace}

\begin{document}
\begin{titlepage}
\pubblock

\vfill
\Title{First experimental investigation of the running \\ of the top quark mass}
\vfill
\Author{ Matteo M. Defranchis \\ on behalf of the CMS Collaboration} 
\Address{\institute}
\vfill
\begin{Abstract}
The first experimental investigation of the running of the top quark mass  is presented. The running, defined in the modified minimal subtraction (\msbar) renormalization scheme, is extracted from a measurement of the differential top quark-antiquark~(\ttbar) production cross section as a function of the invariant mass of the \ttbar system using theoretical predictions at next-to-leading order. The measurement is performed using proton-proton collisions at a centre-of-mass energy of 13\TeV recorded by the CMS detector at the CERN LHC in 2016, corresponding to an integrated luminosity of 35.9\fbinv. Candidate \ttbar events are selected in the final state with an electron and a muon of opposite charge, and the differential cross section is determined at the parton level by means of a maximum-likelihood fit to multidifferential distributions. The extracted running is found to be compatible with the solution of the corresponding renormalization group equation, up to a scale on the order of 1\TeV.

\end{Abstract}
\vfill
\begin{Presented}
$12^\mathrm{th}$ International Workshop on Top Quark Physics\\
Beijing, China, September 22--27, 2019
\end{Presented}
\vfill
\end{titlepage}
\def\thefootnote{\fnsymbol{footnote}}
\setcounter{footnote}{0}

\section{Introduction}
\label{sec:intro}

Beyond leading order in perturbation theory, the parameters of the quantum chromodynamics (QCD) Lagrangian of  undergo renormalization. In the modified minimal subtraction (\msbar) renormalization scheme, the strong coupling constant \as and the top quark masses depend on the renormalization scale. This dependence is described by the renormalization group equations (RGEs), which in the case of the quark masses can be written as:
\begin{equation}
\mu^2 \frac{\mathrm{d} m(\mu)}{\mathrm{d} \mu^2} = - \gamma(\as(\mu))\,m(\mu),
\label{eq:RGE}
\end{equation}
where $\gamma(\as(\mu))$ is the quark mass anomalous dimension and $m(\mu)$ is the quark mass evaluated at the scale $\mu$. Eq.~(\ref{eq:RGE}) can be solved in perturbation theory by expanding the anomalous dimension in powers of \as. Measuring the running of the QCD parameters therefore provides a test of the validity of perturbative QCD. 

The running of \as has been verified on a wide energy range using data from different experiments. Measurements have also been performed to investigate the running of the b~quark and charm quark masses, while the running of the top quark mass~(\mt) has never been probed before. In this analysis, the running of \mt is extracted by comparing next-to-leading order (NLO) theoretical predictions to a measurement of the differential top quark-antiquark (\ttbar) production cross section as a function of the invariant mass of the \ttbar system, \mtt. The analysis is described in detail in Ref.~\cite{Sirunyan:2019jyn}, and the most relevant aspects and results are presented here.

\section{Measurement of the differential cross section}
\label{sec:xsec}

The differential \ttbar production cross section \dstt is determined at the parton level by means of a maximum-likelihood fit to multidifferential distributions of final state observables. In the fit, the systematic uncertainties are treated as nuisance parameters and constrained within the visible phase space. The measurement is performed using proton-proton collisions at a centre-of-mass energy of 13\TeV recorded by the CMS detector at the CERN LHC in 2016, corresponding to an integrated luminosity of 35.9\fbinv. Candidate \ttbar events are selected in the final state with an electron and a muon of opposite charge. In particular, the leading (subleading) lepton is required to have a transverse momentum $\pt > 25\, (20) \GeV$. Furthermore, jets with $\pt > 30\GeV$ are considered, but no selection on the jet multiplicity is applied. In events with at least two jets, the kinematic properties of the top quark and antiquark are estimated by means of an analytical kinematic reconstruction algorithm. This allows to estimate the invariant mass of the \ttbar system, \mttreco.
The dependence on the value of the top quark mass assumed in the kinematic reconstruction and in the Monte Carlo~(MC) simulation, \mtmc, is taken into account in the fit.

In order to obtain unfolded results, the \ttbar MC sample is split into subsamples in intervals of parton-level \mtt corresponding to the desired binning in \dstt. Each subsample is then treated as an independent signal in the fit. A representative scale~\muk is also assigned to each signal, which is chosen to be the centre-of-gravity of bin~$k$ in~\mtt. However, the final result does not depend on the exact definition of~\muk. By expressing the likelihood as a function of the different signal contributions, the results are automatically unfolded to the parton level. This method allows the systematic uncertainties to be constrained simultaneously with \dstt.

The fit is performed in categories of b-tagged jet multiplicity. Events with at least two jets are further divided in subcategories of \mttreco, while other events are categorized separately. A suitable final state distribution is then chosen for each category, as described in Ref.~\cite{Sirunyan:2019jyn}. In particular, the \pt of the softest jet is used to constrain the jet energy scales, while the \mlb distribution, defined as the minimum invariant mass obtained when combining a lepton and a b-tagged jet, is used to determine \mtmc.

\begin{figure}
	\includegraphics[width=.5\textwidth]{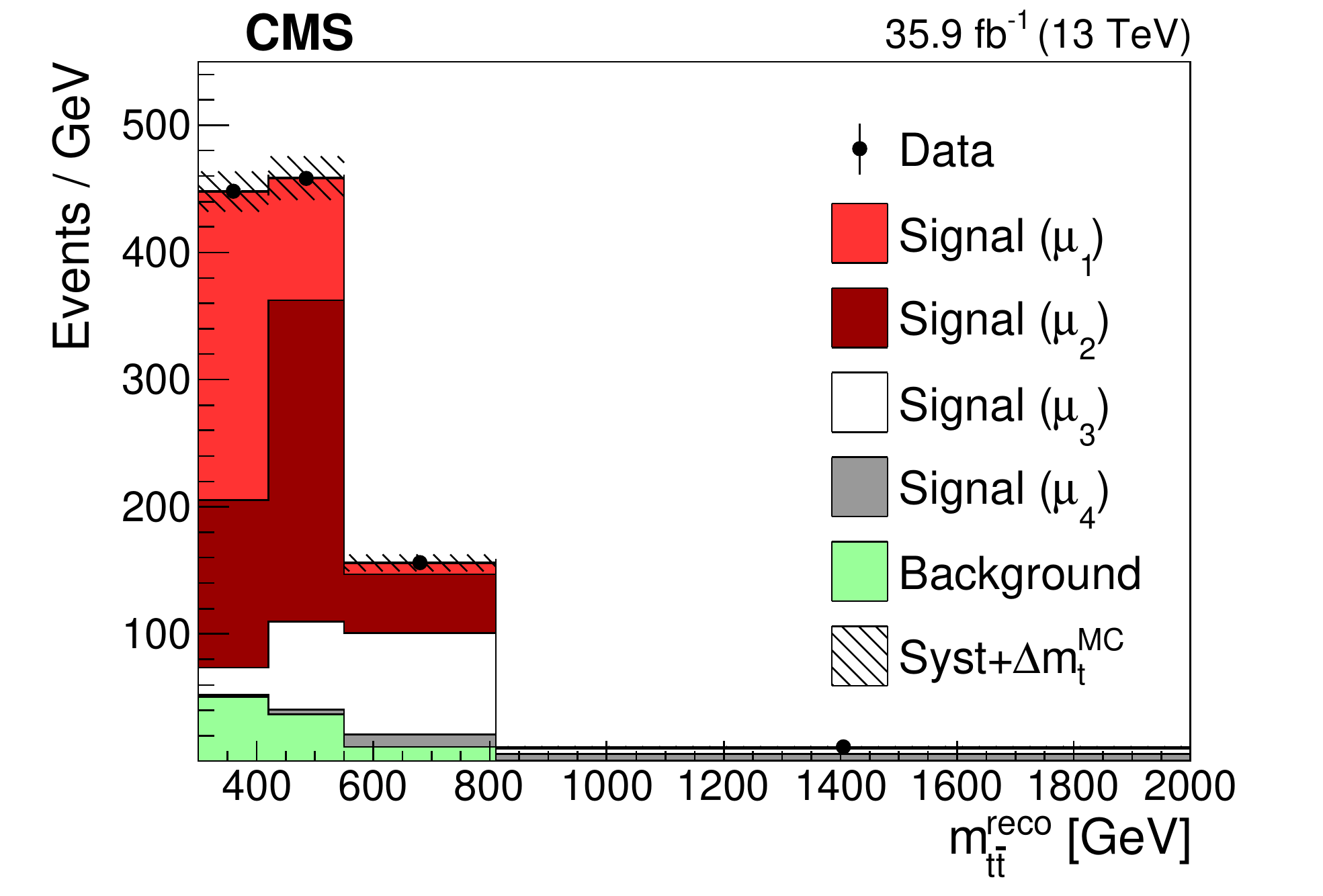}
	\includegraphics[width=.5\textwidth]{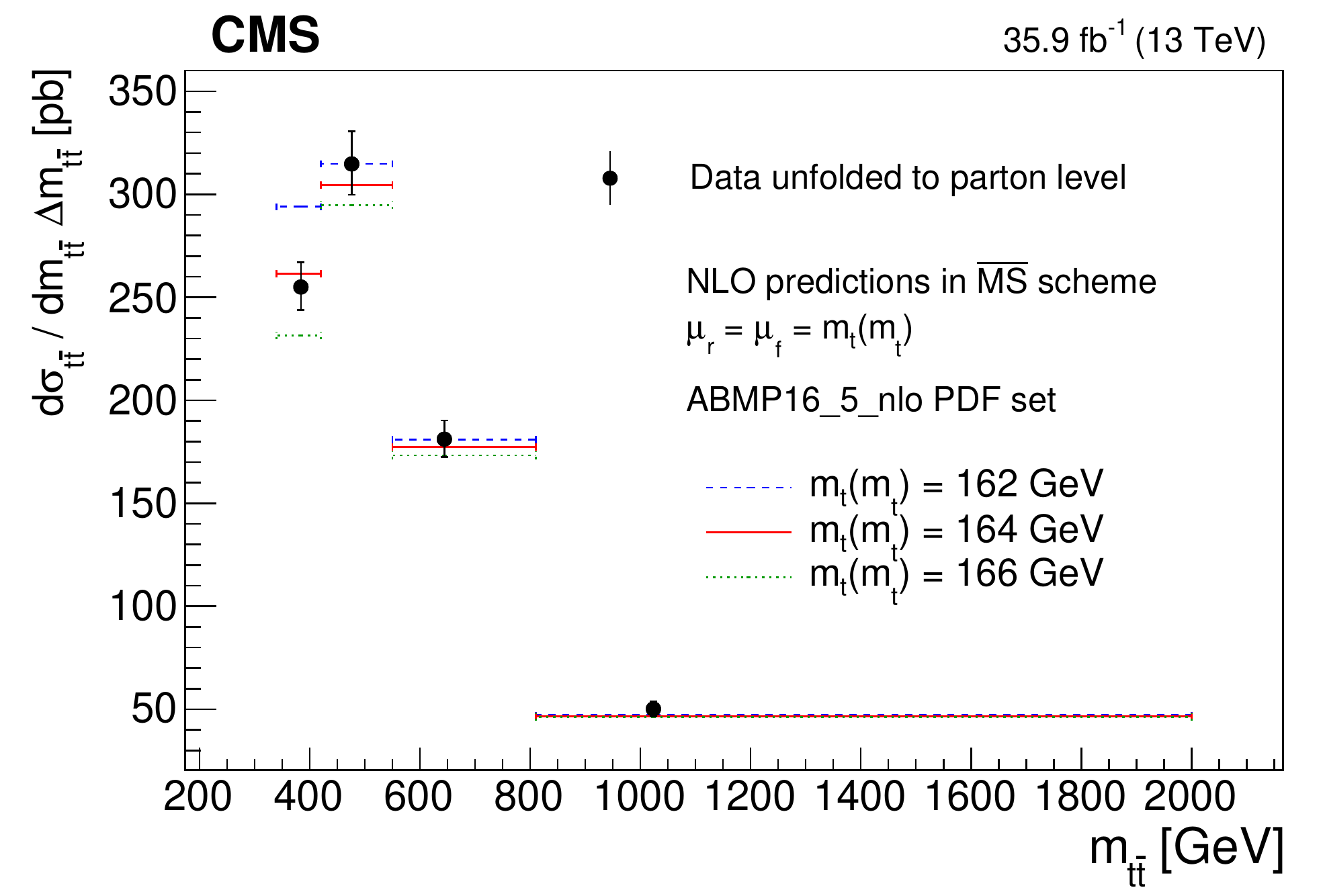}
	\caption{Distribution of \mttreco after the fit to the data (left), and measured \dstt compared to theoretical predictions obtained with different values of \mtmt (right)~\cite{Sirunyan:2019jyn}.
		\label{fig:mtt_xsec}}
\end{figure}

The distribution of \mttreco after the fit can is shown in Figure~\ref{fig:mtt_xsec}~(left), and good agreement between data and the simulation is obtained. In order to ensure that the constraints on the modelling uncertainties are not propagated outside the visible phase space, additional uncertainties are assigned to the extrapolation procedure. In Figure~\ref{fig:mtt_xsec}~(right), the measured \dstt is compared to NLO theoretical predictions obtained with a version of \MCFM where the quark masses are treated in the \msbar scheme~\cite{Dowling:2013baa}. In the calculation, the renormalization and factorization scales are both set to \mt. The predictions are obtained assuming different values of \mtmt, where \mtmt denotes the \msbar mass of the top quark evaluated at the scale $\mu = \mt$. The calculation is interfaced with the ABMP16 NLO set of parton distribution functions~(PDFs)~\cite{Alekhin:2018pai}. As expected, the region close to the \ttbar production threshold provides the highest sensitivity to \mt.

\section{Extraction of the running of the top quark mass}
\label{sec:running}

The measured \dstt and the corresponding theoretical predictions presented in the previous section are used to extract the value of \mtmt independently in each bin of \mtt by means of a \chisq fit of the predictions to the data. The extracted values of \mtmt are then converted to the corresponding \mtmuk using solutions of Eq.~(\ref{eq:RGE}) at one-loop precision, assuming five active flavours. The \chisq definition incorporates the uncertainties in the measured cross section in the visible phase space, while extrapolation and PDF uncertainties are treated separately.

In order to benefit from the partial cancellation of systematic uncertainties, the running is determined with respect to a reference scale \muref, which is chosen to be $\mu_2 = 476\GeV$. However, the final result does not depend on the choice of the reference scale. From the experimental point of view, the quantities $r_{k2} = \mtmuk/\mt(\mu_2)$ are determined, and the results are compared to the predicted running $r(\mu) = \mtmu/\mt(\muref)$, which solely depends on the solution of the RGE. Variations of the renormalization and factorization scales are not performed, as the RGE is probed at a fixed order in perturbation theory. When evaluating the ratios, all correlations are properly taken into account. 

\begin{figure}
\includegraphics[width=.5\textwidth]{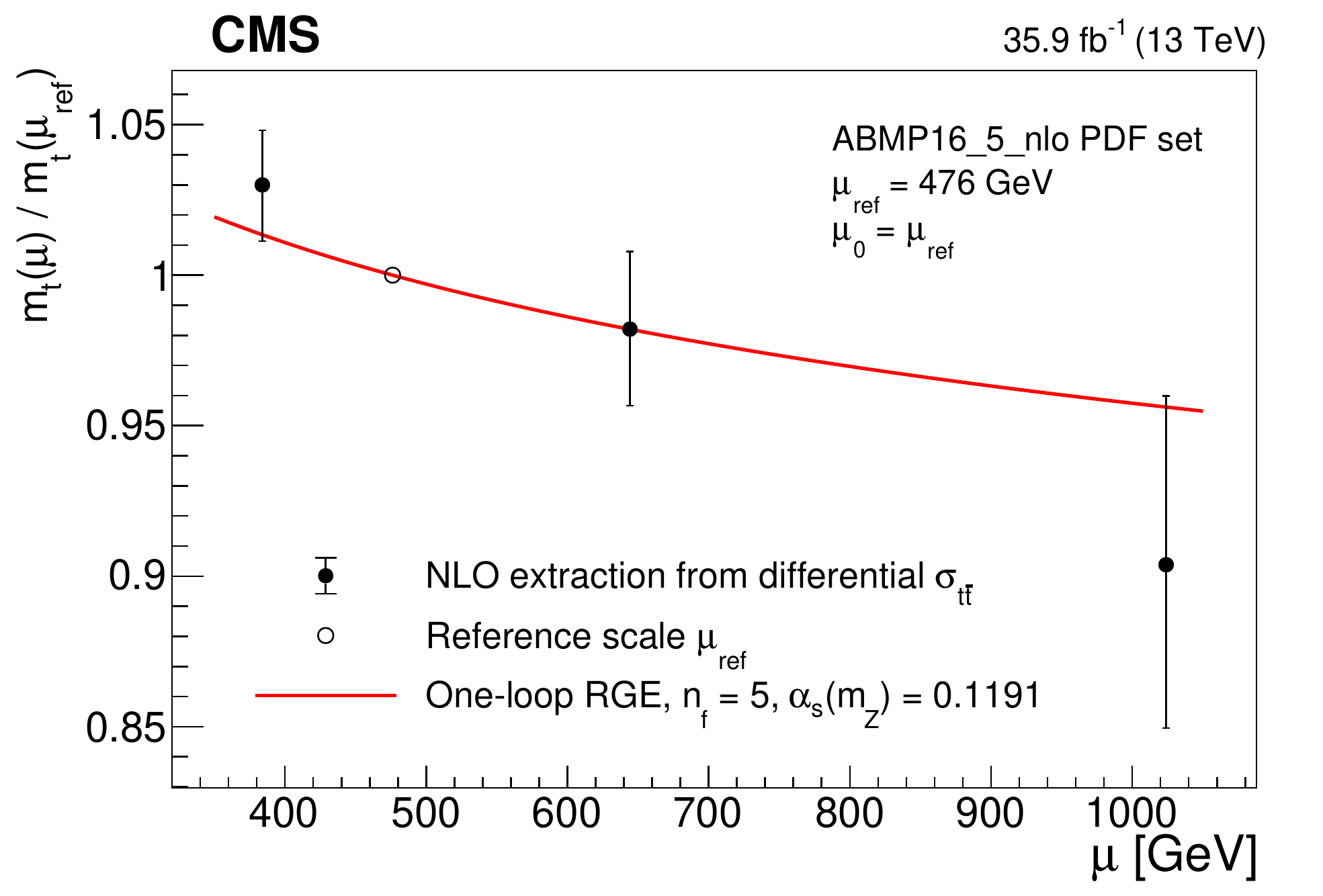}
\includegraphics[width=.5\textwidth]{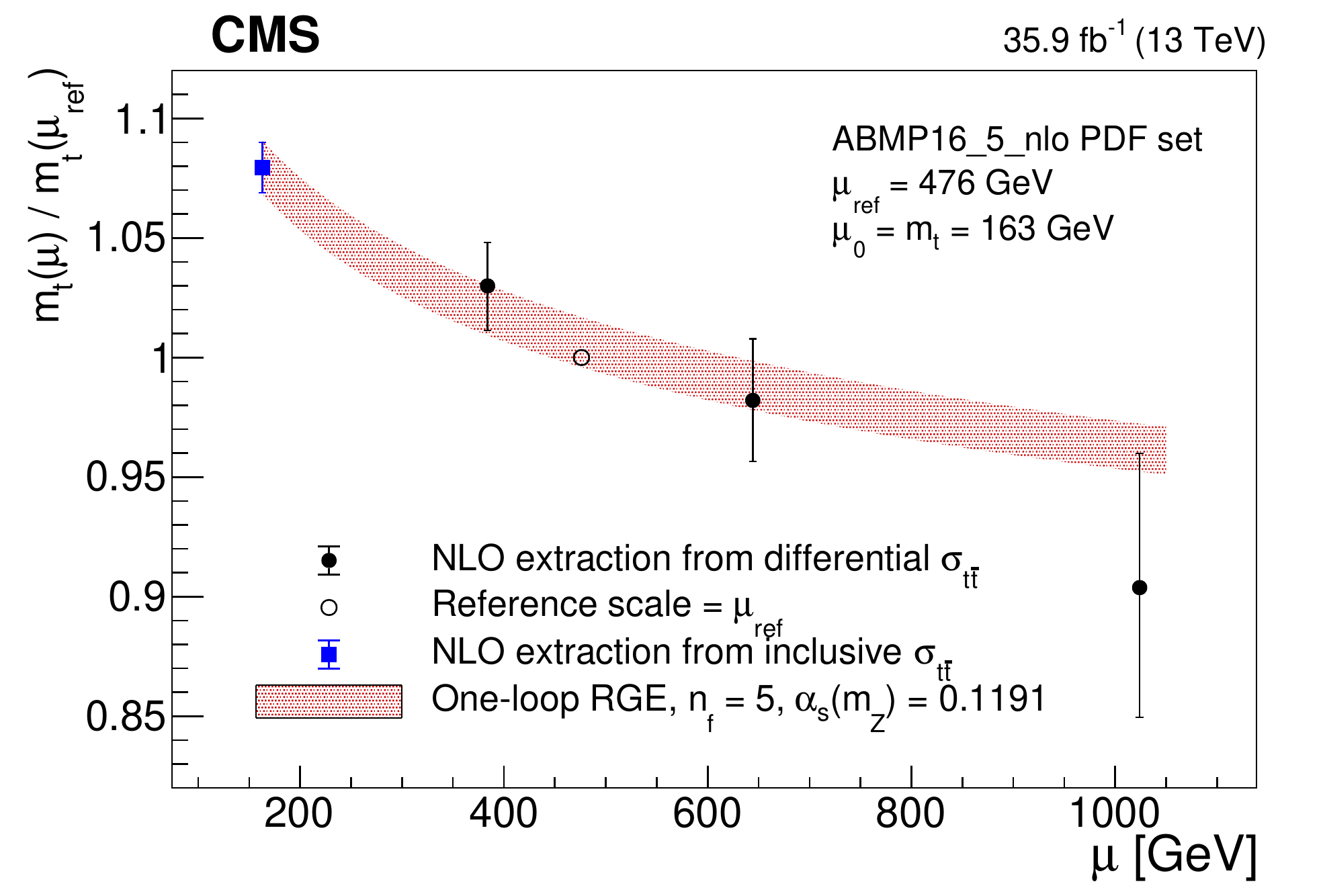}
\caption{Left: extracted running of the top quark mass compared to the RGE solution at one-loop precision evolved from the initial scale $\mu_0 = 476\GeV$. Right: comparison of the result with the value of \mtot, \ie the value of \mtmt extracted at NLO from the measurement of the inclusive \ttbar cross section of Ref.~\cite{Sirunyan:2018goh}. The uncertainty in \mtot is evolved from the initial scale $\mu_0 = \mt = 163\GeV$ using the same RGE solution~\cite{Sirunyan:2019jyn}.
\label{fig:running}}
\end{figure}

The extracted running is shown in Figure~\ref{fig:running}~(left), and good agreement with the one-loop RGE solution is observed in the investigated range. In Figure~\ref{fig:running}~(right), the result is compared to $\mtot/\mt(\mu_2)$, where \mtot is the value of \mtmt extracted at NLO from the inclusive \ttbar cross section (\stt) measured in Ref.~\cite{Sirunyan:2018goh}. The uncertainty in \mtot includes the uncertainty in \stt and the PDFs, while the value of $\mt(\mu_2)$ in the ratio $\mtot/\mt(\mu_2)$ is taken without uncertainty.

Finally, the compatibility between the extracted running and RGE solution is quantified by parameterizing the measured ratios with the function
\begin{equation}
f(x,\mu) = x\left[ r(\mu) -1 \right] +1 ,
\label{eq:fit_sign}
\end{equation}
which corresponds to the RGE solution for $x=1$, and to a hypothetical no-running scenario for $x=0$. The best-fit value of $x$, denoted with $\hat{x}$, is determined by means of a \chisq fit to the measured ratios, taking all correlations into account. The result is:
\begin{equation}
  	\hat{x} =  2.05 \pm 0.61~ (\mathrm{fit}) ~^{+0.31}_{-0.55}~ (\mathrm{PDF}+\as) ~ ^{+0.24}_{-0.49}~ (\mathrm{extr}) .
\end{equation}
Here, the uncertainties denote the uncertainty in the visible cross section (fit), the combined PDF and \as uncertainty as determined using the ABMP16 eigenvectors, and the extrapolation uncertainty (extr). Based on this result, the extracted running is found to be compatible with the one-loop RGE solution within 1.1 standard deviations, while the no-running hypothesis is excluded at above 95\% confidence level.

The precision of the analysis is limited by the uncertainties in the integrated luminosity, the lepton identification efficiencies, the jet energy scales, and the modelling of the \ttbar simulation, while the statistical uncertainty is found to be negligible.

\Acknowledgements
The author would like to thank the PIER Helmholtz Graduate School for financing his participation to the conference.

\end{document}